\documentclass{emulateapj}

\usepackage{graphicx}
\usepackage{subfigure}
\usepackage{latexsym}

\newcommand{\HI}{\protect\ion{H}{1}}
\newcommand{\msun}{$M_\odot$}
\newcommand{\etal}{{et~al.}}
\newcommand{\mhi}{$M_{HI}$}
\newcommand{\kms}{~km\,s$^{-1}$}

\begin{document}
\title{Where are the High Velocity Clouds in Local Group Analogs?}
\author{D.J. Pisano\altaffilmark{1,2,3}, David G. Barnes\altaffilmark{4,5}, 
Brad K. Gibson\altaffilmark{6,7}, Lister Staveley-Smith\altaffilmark{1,8},\\
Ken C. Freeman\altaffilmark{9,10}, Virginia A. Kilborn\altaffilmark{6,11}}

\altaffiltext{1}{Australia Telescope National Facility, P.O. Box 76, Epping NSW 1710, 
Australia}
\altaffiltext{2}{DJ.Pisano@atnf.csiro.au}
\altaffiltext{3}{NSF MPS Distinguished International Postdoctoral Research 
Fellow, Bolton Fellow}
\altaffiltext{4}{School of Physics, Univ. of Melbourne, Victoria 3010, Australia}
\altaffiltext{5}{dbarnes@astro.ph.unimelb.edu.au}
\altaffiltext{6}{Centre for Astrophysics \& Supercomputing, Swinburne University, 
Hawthorn, Victoria 3122, Australia}
\altaffiltext{7}{bgibson@swin.edu.au}
\altaffiltext{8}{Lister.Staveley-Smith@atnf.csiro.au}
\altaffiltext{9}{RSAA, Mount Stromlo Observatory, Cotter Road, Weston, ACT 2611, 
Australia}
\altaffiltext{10}{kcf@mso.anu.edu.au}
\altaffiltext{11}{vkilborn@astro.swin.edu.au}

\begin{abstract}
High-velocity clouds (HVCs) are clouds of \HI\ seen around the Milky
Way with velocities inconsistent with Galactic rotation, have unknown 
distances and masses and controversial  origins.  One possibility is
that HVCs are associated with the small dark matter halos seen in
models of galaxy formation and distributed at distances of 150 kpc - 1
Mpc.  We report on our attempts to detect the analogs to such putative
extragalactic clouds in three groups of galaxies similar to our own
Local Group using the ATNF Parkes telescope and Compact Array.  Eleven
dwarf galaxies were found, but  no \HI\ clouds lacking stars were
detected.   Using the population of compact HVCs around the Milky Way
as a template, we find that our non-detection of analogs implies that
they must be clustered within 160 kpc of the Milky Way  (and other
galaxies) with an average \HI\ mass $\lesssim$4$\times$10$^5$\msun\ at
the 95\% confidence level.  This is in accordance with recent
limits  derived by other authors.  If our groups are true analogs to the
Local Group, then this makes the original Blitz \etal\ and Braun \& Burton
picture of HVCs residing out to 1 Mpc from the Milky Way extremely unlikely.  
The total \HI\ mass in HVCs, $\lesssim$1$\times$10$^8$\msun, implies that 
there is not a large reservoir of neutral hydrogen waiting to be accreted
onto the Milky Way.  Any substantial reservoir of baryonic matter must be 
mostly ionized or condensed enough as to be undetectable.  
\end{abstract}


\section{Introduction}
\label{intro}

Forty years after first being discovered (Muller, Oort, \& Raimond 1963), the 
``high-velocity clouds'' (HVCs) remain a mystery.  HVCs are clouds of neutral
hydrogen (\HI) covering a large fraction of the entire sky with velocities 
inconsistent with simple Galactic rotation and in excess of $\pm$90 \kms\ of the
Local Standard of Rest (see Wakker \& van Woerden 1997 for a review).  In addition, 
they lack stellar emission (e.g. Simon \& Blitz 2002).  These facts 
make the determination of distances an intractable problem;  without distances
we are unable to determine their masses and discriminate between mechanisms 
responsible for their origins.

HVCs most likely represent a variety of phenomena.  Some HVCs are probably
related to a galactic fountain (Shapiro \& Field 1976; Bregman 1980) and are
located in the lower Galactic halo.  Other HVCs are certainly tidal in origin:
the Magellanic Stream is the most obvious of these features, formed via the 
tidal interactions between the Milky Way, Large Magellanic Cloud, and Small
Magellanic Cloud (e.g. Putman \etal\ 1998), with other HVCs potentially related 
to the Sagittarius dwarf (Putman \etal\ 2004).  Some HVCs may even be satellites
unto themselves (e.g. Lockman 2003).  Oort (1966, 1970) originally proposed that 
HVCs may be infalling primordial gas; Complex C may be such an example 
(Wakker \etal\ 1999; Tripp \etal\ 2003; cf. Gibson et al. 2001).   Verschuur (1969) 
was the first to associate HVCs
with the Local Group, with the idea resurrected by Blitz \etal\ (1999) for
all HVCs and by Braun \& Burton (1999) for the subset of compact HVCs (CHVCs).  
These authors suggested that HVCs contained dark matter and could be related 
to the small dark matter halos predicted to exist in large numbers by cold dark
matter models of galaxy formation (e.g. Klypin \etal\ 1999, Moore \etal\ 1999).
In this scenario, Blitz \etal\ and Braun \& Burton hypothesize that HVCs have D$\sim$1 
Mpc, and \mhi $\sim$10$^7$\msun.  Since these papers, much of the observational
effort has focused on testing the association of CHVCs with dark matter halos 
and the formation of the Local Group.
In addition, distance and mass estimates have decreased; de Heij \etal\ (2002b)
suggested the CHVC distribution has a Gaussian distribution about the 
Milky Way and M31 with D$\sim$150-200 kpc and \mhi $\sim$10$^{5.5-7}$\msun, but are
still associated with dark matter halos.  

If the Blitz \etal\ (1999), Braun \& Burton (1999), and de Heij \etal\ (2002b) 
hypothesis is correct, then analogs to HVCs should be ubiquitous in other galaxy
groups.  Numerous attempts to find extragalactic analogs to HVCs have been initiated,
but, to date, there have been no discoveries.  A few authors have reported high 
velocity gas around individual galaxies, but these HVCs are probably associated 
either with vigorous star formation (e.g. Schulman \etal\ 1994, Kamphuis \& 
Sancisi 1993) or with tidal interactions (e.g. Kamphuis \& Briggs 1992).  Pisano, 
Wilcots, \& Liu (2002) searched for \HI\ clouds around 41 isolated, quiescent 
galaxies. While discovering 13 companions, all were dwarf galaxies.  These
studies all assumed that HVCs were associated with individual galaxies, 
while they may instead be unique to the group environment.  

Lo \& Sargent (1979) conducted one
of the earliest searches for intergalactic HI in three loose groups.  They
detected four dwarf galaxies, but lacked the sensitivity to detect HVC 
analogs.  Over the typical FWHM of a CHVC ($\sim$30 \kms) their 5$\sigma$ 
sensitivity was 4$\times$10$^7$ - 5$\times$10$^8$\msun.  A more recent
survey of one of the same groups, Canes Venatici I, by Kraan-Korteweg 
\etal\ (1999) had a detection limit of $\sim$10$^8$\msun and also failed
to find anything more than typical dwarf galaxies.  Other studies of groups, such 
as those by Zwaan \& Briggs (2000), Zwaan (2001), and de Blok \etal\ (2002),
only probed a small fraction of their total area reducing the probability
of detecting analogs.  In addition, these surveys did not explore spiral-rich groups
akin to the Local Group.  The most sensitive group survey to date is the Parkes HIDEEP
survey (Minchin \etal\ 2003) which covered part of the Cen A group.  Despite
their 5$\sigma$ \mhi\ detection limit of 2$\times$10$^6$\msun\ over 30 \kms\ at
the distance of Cen A (= 3.5 Mpc; Cot\^e \etal\ 1997), Minchin
\etal\ found no sources without optical counterparts; i.e. no HVC analogs.

Despite the large number of searches for HVCs, all these studies have crucial
limitations.  Some lack the sensitivity to detect HVC analogs and others only 
surveyed a small region of the group reducing the number of expected 
detections.  Perhaps most critically, however, is the lack of observations of 
groups like the Local Group.  The Cen A group, for example, is a fairly dense 
group centered around a large elliptical galaxy, quite unlike the Local Group. 
If HVCs are unique to the relatively tame environment of the Local Group, then 
we may not expect to see them in the Cen A group or groups like it.  

In this Letter, we present our observations of three loose groups of galaxies 
analogous to the Local Group with the Parkes multibeam and Australia Telescope 
Compact Array (ATCA)\footnote{The Parkes telescope and ATCA are part of the 
Australia Telescope which is funded by the Commonwealth of Australia for operation
as a National Facility managed by CSIRO} and discuss the implications for the 
location of HVCs around the Milky Way.  In Section~\ref{obs} we 
discuss the group properties and our observations.  In Section~\ref{model}, we 
describe a model for the distribution of HVCs in the Local Group and its predictions
for what we should see in our sample of groups.  Finally, we conclude in Section~\ref{disc} 
by comparing our observations with the model prediction and what this implies 
for HVCs in the Local Group.

\section{Observations}
\label{obs}

We chose to observe three loose groups of galaxies which were qualitatively 
similar to the Local Group:  LGG 93, LGG 180, and LGG 478 (Garcia
1993).  These groups were selected to contain only large spiral galaxies 
which were typically separated by $\gtrsim$100 kpc and spread over a 
diameter of $\sim$1 Mpc.  All of the groups are nearby, between 10.6 and 
13.4 Mpc.  At this distance, the Parkes beam of 14.4$^\prime\ $ corresponds to
a linear size of $\sim$45 kpc.  Between October 2001 and August 2002, we observed 
an area of $\sim$1 Mpc$^2 \equiv$25 square degrees over a velocity range of $>$ 1500 
\kms\ centered on each group by scanning the Multibeam receiver on the Parkes 
telescope down to an rms sensitivity of 6-9 mJy beam$^{-1}$ per channel.  
This translates to an rms mass sensitivity of $\lesssim$10$^6$\msun\ per 3.3 \kms.  
Fake sources were inserted into the cubes, and multiple double-blind searches 
for all real and fake sources were conducted.  All sources, not just new ones, 
identified by more than one search were confirmed with follow-up observations 
using the ATCA.  Based on our identification 
of the fake sources, we determined that we detected all sources which had an 
integrated flux greater than 10 times the rms noise times the square root of the 
number of channels.  This means that over a velocity width of $\sim$35 \kms\ 
(the average FWHM velocity width of CHVCs), we can only detect sources in our
Parkes and ATCA data with \mhi\ $\gtrsim$10$^7$\msun.  More detail on the groups, 
observations, reductions, and analysis will be presented in Pisano \etal\ (2004, 
in preparation).  The properties of the groups and the observations are listed in 
Table~\ref{groups}.

In the three groups, all the known members were detected and eight new
\HI-rich dwarf galaxies were found with optical counterparts visible on the Digital 
Sky Survey or cataloged in NED; no \HI\ clouds without stars 
were discovered.  In other words, no HVC analogs were found with \mhi\ 
$\gtrsim$10$^7$\msun.  At the distance of these groups, the Parkes
beam is 45 - 55 kpc, but since our higher spatial resolution ATCA 
observations have the same sensitivity as the Parkes data we should detect 
any massive HVC analog that is more than $\sim$ 5 kpc from a galaxy.

\section{A Model for CHVCs}
\label{model}

Because we did not detect any HVC analogs in the three groups surveyed, we are
unable to directly measure the masses and spatial distributions of such clouds.
Since these three groups are similar to the Local Group in terms of 
their morphology and the \HI\ and halo mass functions (Pisano \etal\ 2004, in
preparation), we can use our non-detections to infer the distribution of HVCs
within the Local Group.  To this end, we have constructed a simple model for 
the distribution of CHVCs around the Milky Way and other galaxies.  Because 
CHVCs are the most likely class of HVCs to be dark matter dominated and reside 
at larger distances from the Milky Way, we only consider this class of objects 
(Braun \& Burton 2001).  

\begin{figure}
\plottwo{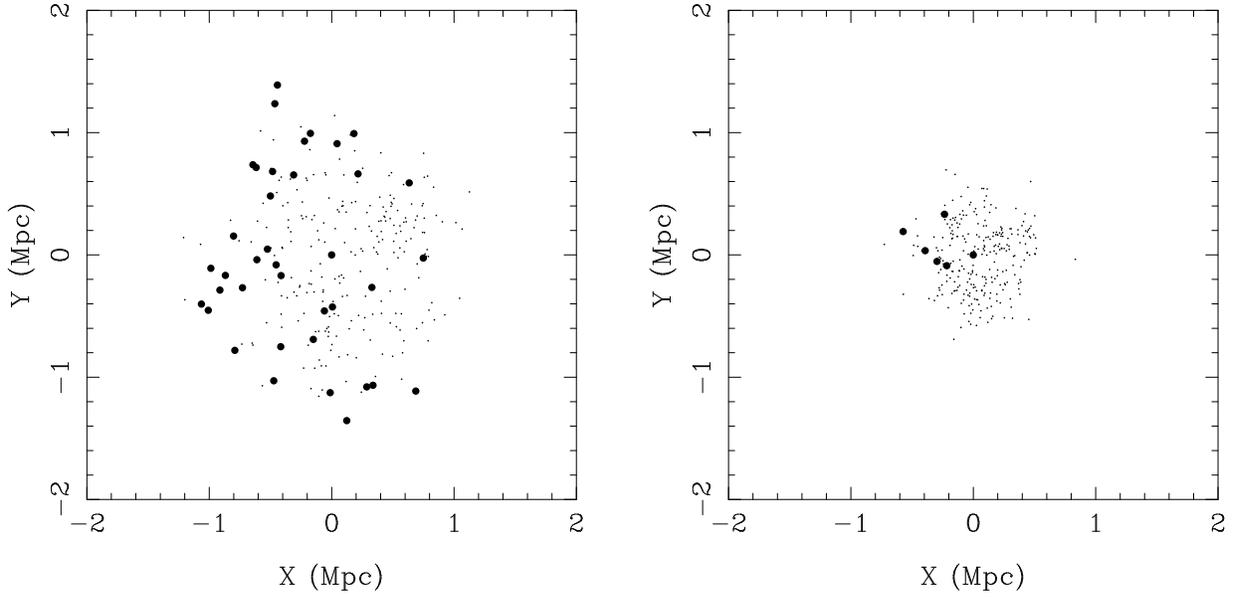}{figure1b.ps}
\caption{Left (a):  A simulation of all 270 cataloged CHVCs from Putman \etal\ 
(2002) and de Heij \etal\ (2002a) around the Milky Way (large solid circle in 
center) distributed with a random three-dimensional Gaussian distance distribution 
with D$_{HWHM}$ = 500 kpc.  Solid circles would be detected by our survey, 
dots would not if the Milky Way were at the distance of our groups.  
Right (b): Same as (a), but for D$_{HWHM}$ = 250 kpc.  Note 
that there are many fewer expected detections.\label{fig:model}}  
\end{figure}

For our model, we start with the cataloged CHVCs from Putman \etal\ (2002)
and de Heij \etal\ (2002a) from the southern HIPASS and northern Leiden-Dwingeloo surveys.
This yields 270 CHVCs with measured fluxes and velocity widths.  We 
assume these clouds are distributed with a three-dimensional Gaussian distance 
distribution around the Milky Way characterized by a given D$_{HWHM}$.  
After assigning a distance to a cloud, we get an \HI\ mass which we compare to 
our 10$\sigma$ detection limit for the cloud's velocity width to determine if 
we could detect this cloud in one of our groups.  We carry out a Monte Carlo
simulation with 10,000 trials noting the number of times we have zero detections.  
We do this for a variety of D$_{HWHM}$ values, ranging from 50 kpc to 500 kpc, 
and for differing parent numbers of CHVCs, ranging from 27 to 1728 clouds (0.1 - 
6.4 times the number of Galactic CHVCs).  Two examples of this model are presented 
in Figure~\ref{fig:model} for a total number of 270 CHVCs with D$_{HWHM}$ = 500 kpc 
(left) and 250 kpc (right).  Note that for the latter model, distinctly fewer CHVCs 
would have been detected in our survey.  While our model is distinctly less complex 
than those of previous authors (e.g. Blitz \etal\ 1999, Braun \& Burton 1999, de Heij
\etal\ 2002b) that include assumptions as to the physical properties of HVCs, it can 
be seen as a generalization of these models.  For reference, 
the Blitz \etal\ (1999) and Braun \& Burton (1999) models have D$_{HWHM} 
=$ 500 kpc while the de Heij \etal\ (2002b) model has D$_{HWHM} =$ 150 kpc.  

There are a few important aspects of this model which may limit its potential 
utility.  First of all, as can be seen in Figure~\ref{fig:model}, we do not
expect to detect the vast majority of CHVCs at the distance of our groups, 
but only the most massive.  As D$_{HWHM}$ decreases, this becomes more of an
issue.  For example, at D$_{HWHM}$ = 500 kpc, the average CHVC 
\HI\ mass is $\sim$10$^7$\msun, while our detection limit is over 10$^7$\msun, 
but at D$_{HWHM}$ = 250 kpc, the average \HI\ mass is only $\sim$10$^6$\msun.   
As such, the detailed nature of the flux and linewidth distributions of CHVCs 
around the Milky Way is of critical importance.  If this is different around 
other galaxies in other groups, in particular if the highest flux CHVCs are 
absent in such groups, then this model may not yield accurate limits. 

It is also important to note that the number of CHVCs observed around the 
Milky Way may not be equal to the total number present, neither of which need
be equal to the number around galaxies in other groups.  This is why we vary
the total number of CHVCs in our model.  If the number is higher, then the
constraints will be stronger.  If other types of HVCs are considered or the existing
catalogs of CHVCs are incomplete, then, again, we would expect to detect more 
analogs so our distance constraints would be more stringent.   We can, however, 
make a rough estimate of how many
clouds we expect in each group.  Cold dark matter (CDM) simulations of the
formation of the Local Group (Klypin \etal\ 1999), show that the number of 
satellites per galaxy is proportional to the mass of that galaxy, which is 
proportional to the cube of that galaxy's circular velocity, 
$N_{CHVC}\propto M_{galaxy} \propto V_{circ}^3$.  This can also be argued via the
Tully-Fisher relation (Tully \& Fisher 1977).  Using published inclinations,
and measured velocity width for each group galaxy, the number of expected
CHVCs in each group is within a factor of two of the number seen around the Milky 
Way.  This is accounted for in our model comparisons, but will not have a major effect 
on our distance limits.  Finally, it is possible that HVCs are present in all of 
the groups we observed, but that they effectively cover the entire area of the 
group.  In this case, in our reductions, we would have subtracted out the real 
signal as sky.  This is unlikely as Milky Way HVCs only cover 37\% of the sky down
to a column density of 7$\times$10$^{17}$cm$^{-2}$ (Murphy, Lockman, \& Savage 1995).
Furthermore, such a distribution would be inconsistent with the statistics of {\sc MgII}
and Lyman limit absorption line systems seen towards quasars (Charlton, Churchill, \&
Rigby 2000).  

\section{Where are the CHVCs?}
\label{disc}

\begin{figure}
\plotone{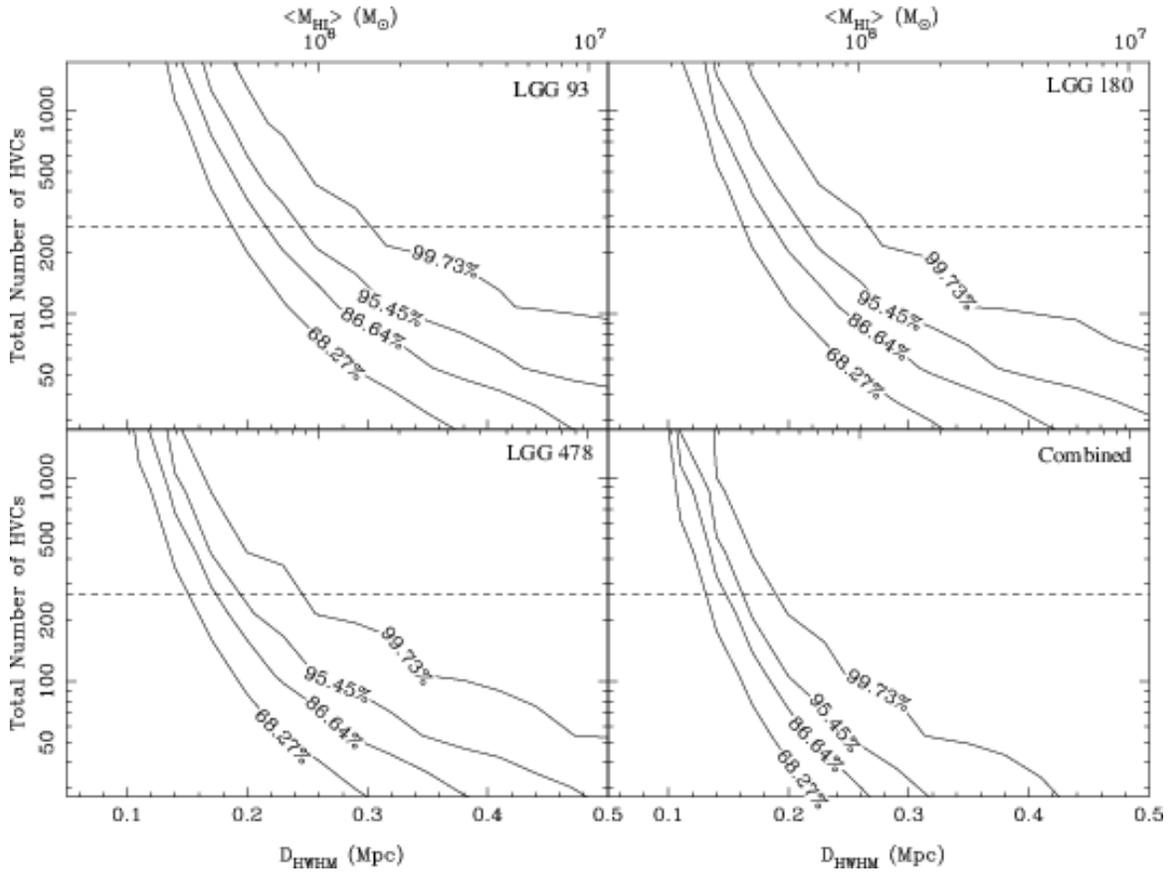}
\caption{A plot of the probability of zero detections as a function of the
number of CHVCs per group and D$_{HWHM}$ (or the average \HI\ mass
of the CHVC) for the distribution of Milky Way CHVCs for each group and the 
combined probability for all three groups as labeled on the panels.  The dashed 
line marks the number of CHVCs identified around the Milky Way.\label{fig:results}}
\end{figure}

Figure~\ref{fig:results} shows the probability of zero detections as a
function of the parent number of CHVCs, D$_{HWHM}$, and the average
\mhi\ of CHVCs for each group and the combined probability for the
three groups.  We  can combine the individual group probabilities
since they are independent experiments.  The figure shows that our
non-detection of HVC analogs means  that the average \HI\ mass of
CHVCs must be less than 10$^6$\msun at the 95.45\%  confidence level.
This assumes that the properties of
CHVCs in these groups  are the same as those in cataloged in the Local
Group.  If this is the case, then we can infer that for this \HI\
mass, CHVCs in the Local Group must be clustered within D$_{HWHM} <$
160 kpc of the Milky Way.  If we were to consider all HVCs in our
model, then these limits would be even stronger.  This conclusion is
robust even when considering different models for the CHVC
distribution.  The average \HI\ mass of CHVCs  is the same if they are
distributed in a filamentary manner or if we adopt the de Heij \etal\
(2002b) model.  These limits are inconsistent with original models of
Blitz \etal\ (1999) and  Braun \& Burton (1999) which would have 
median distances of $\sim$1 Mpc with an \mhi\ of $\sim$10$^7$\msun.  
Our sensitivity is not sufficient to constrain the best fit de Heij \etal\ (2002b) 
model.

Our results, when compared with those of other authors, reveal a remarkably 
consistent picture for the distribution of HVCs in close proximity to individual
galaxies.  Zwaan's (2001) study of parts of five groups with Arecibo constrained HVCs 
to be within 200 kpc of group barycenters.  Braun \& Thilker (2004) 
and Thilker \etal\ (2004) report on a possible population of HVCs around M31 
with \mhi\ ranging from 10$^{5-7}$\msun\ and a Gaussian distance dispersion of 
55 kpc.  Attempts to measure or model the distances to HVCs observed around the
Milky Way also point to this same picture.  The few direct stellar absorption line 
distances available for HVC complexes place these clouds within $\sim$10 kpc of 
the Milky Way (Wakker \etal\ 2001).  
Putman \etal's (2003) H$\alpha$ observations of HVCs and CHVCs around the Milky 
Way constrain those clouds to be within $\sim$40 kpc of the Galaxy assuming a
model for the escaping ionizing radiation.  Maloney \& Putman (2002) and 
Sternberg, McKee, \& Wolfire (2002) modeled CHVCs as gaseous objects within 
dark matter halos while accounting for the effects of ionization, thermal
balance and confinement by an external medium and determined that CHVCs must
lie within 150 kpc of the Milky Way.    
Finally, de Heij \etal's (2002b) model of the Local Group distribution of CHVCs 
using their assumed physical properties predicts a distribution with 
D$_{HWHM}\sim$150-200 kpc.

At these distances CHVCs are more closely associated with the Milky Way than 
the Local Group, which suggests that these clouds are associated more with
individual galaxy formation instead of group formation as originally suggested by
Blitz \etal\ (1999).  Also at these distances, the total \HI\ mass in CHVCs is 
$\lesssim$10$^8$\msun, and even with substantial dark matter would only contribute 
a small fraction of the total mass of the Local Group.  They would still contribute 
fuel for star formation in the Milky Way, but only as much \HI\ as a single dwarf 
galaxy.  On the other hand, CHVCs may still be the repository for
large amounts of ionized gas (Maloney \& Putman 2002, Sternberg et al. 2002) which
could condense onto the Milky Way.  Interestingly, the similarity of the inferred 
radial distribution of CHVCs with Milky Way satellites and models of galaxy formation 
(Kravstov \etal\ 2004) may actually strengthen the argument that 
CHVCs are associated with low mass dark matter halos.  Future searches for CHVC analogs 
associated with galaxy formation with properties like those inferred by de Heij \etal\ 
(2002b) will be difficult due to their low masses.  It will
also be difficulty to infer the origin of any such analogs.  Within 160 kpc of a galaxy, 
\HI\ associated with galactic fountains and tidal interactions will be prevalent making 
the identification of CHVCs associated with galaxy formation difficult.  Nevertheless, 
if we can find gas clouds associated with galaxy formation it will not only shed light 
on the nature of high velocity clouds, but serve as a valuable check on models of galaxy 
formation.

\acknowledgements

The authors wish to thank the staff at Parkes and the ATCA for their 
assistance with observing.  We thank Warwick Wilson and the ATNF engineering
group for their excellent work in making the 16 MHz filters for these 
observations.  We thank Chris Brook for help with additional observing and 
Tim Connors for help during the cube searching process.  Finally, we thank 
Bob Benjamin for his insightful comments on this paper as its referee.  D.J.P. acknowledges 
generous support from NSF MPS Distinguished International Research Fellowship 
grant AST0104439.  BKG acknowledges the financial support of the Australian
Research Council.

\begin{deluxetable}{lcccc}
\tablecolumns{5}
\tablewidth{0pc}
\tablecaption{Sample Properties\label{groups}}
\tablehead{\colhead{Group} & \colhead{Galaxies\tablenotemark{a}} & 
\colhead{Velocity\tablenotemark{b}} & \colhead{Distance\tablenotemark{c}} & 
\colhead{Sensitivity\tablenotemark{d}} \\
\colhead{} & \colhead{} & \colhead{km s$^{-1}$} & \colhead{Mpc} & 
\colhead{M$_\odot$}}
\startdata
LGG 93  & 5 S        & 750 & 11.5 & 8$\times$10$^5$\\
LGG 180 & 3 S, 6 Irr & 725 & 11.1 & 6$\times$10$^5$ \\
LGG 478 & 3 S, 1 Irr & 692 & 10.6 & 5$\times$10$^5$ \\
\enddata
\tablenotetext{a}{The morphological types of group galaxies:  S = spiral, 
Irr = irregular.}
\tablenotetext{b}{The recession velocity of the group corrected for Virgocentric infall from Garcia (1993).}
\tablenotetext{c}{The distance to the group calculated from the corrected velocity and assumming H$_0$ = 65 km s$^{-1}$ Mpc$^{-1}$}
\tablenotetext{d}{The 1$\sigma$ mass sensitivity in one 3.3 \kms\ channel.}
\end{deluxetable}

\end{document}